\documentclass{article}

\usepackage{amssymb,amsmath,amsfonts,eurosym,geometry,ulem,graphicx,caption,color,setspace,sectsty,comment,footmisc,caption,pdflscape,array,float,multirow,bm}

\usepackage[utf8]{inputenc}
\usepackage[english]{babel}
 
\newtheorem{theorem}{Theorem}[section]

\newtheorem{result}[theorem]{Result}

\date{}

\begin{document}

\title{Staggered Release Policies for COVID-19 Control: Costs and Benefits of Sequentially Relaxing Restrictions by Age} 

%Analysis of Costs and Benefits Related to COVID-19 Restriction Release Policies Staggered by Age

\maketitle

\vspace{-3em}

\centerline{\large Henry Zhao$^{1 {\ast}}$, Zhilan Feng$^{2,3}$, Carlos Castillo-Chavez$^4$, Simon A. Levin$^{5}$
}

\medskip

\centerline{$^{1}$Department of Economics, Princeton University, Princeton, NJ 08544}

\centerline{$^{2}$Department of Mathematics, Purdue University, West Lafayette, IN
47907}

\centerline{$^{3}$Division of Mathematical Sciences, National Science Foundation, Alexandria, VA 22314}

\centerline{$^{4}$ School of Human Evolution and Social Change, MCMSC,} 

\centerline{Arizona State University, Tempe, 
AZ 85281} 

\centerline{$^{5}$ Department of Ecology \& Evolutionary Biology, Princeton University, Princeton, NJ 08544}

\vspace{1.5em}

\centerline{$^\ast$To whom correspondence should be addressed; E-mail:  hz5@princeton.edu}

\vspace{2em}

\hrule

\vspace{0.5em}

\begin{abstract}

Strong social distancing restrictions have been crucial to controlling the COVID-19 outbreak thus far, and the next question is when and how to relax these restrictions. A sequential timing of relaxing restrictions across groups is explored in order to identify policies that simultaneously reduce health risks and economic stagnation relative to current policies. The goal will be to mitigate health risks, particularly among the most fragile sub-populations, while also managing the deleterious effect of restrictions on economic activity. The results of this paper show that a properly constructed sequential release of age-defined subgroups from strict social distancing protocols can lead to lower overall fatality rates than the simultaneous release of all individuals after a lockdown. The optimal release policy, in terms of minimizing overall death rate, must be sequential in nature, and it is important to properly time each step of the staggered release. This model allows for testing of various timing choices for staggered release policies, which can provide insights that may be helpful in the design, testing, and planning of disease management policies for the ongoing COVID-19 pandemic and future outbreaks.
 
\end{abstract}

\vspace{0.75em}

\hrule

\vspace{0.75em}

\section{Introduction}

Social distancing restrictions like the stay-at-home orders issued in the United States are structured as a type of everybody-or-nobody type of policy. However, limiting attention to this specific class of policies ignores the crucial fact that diseases like COVID-19 have differential impacts of infection on various sub-groups of the population. Disease severity and survival rates for COVID-19 have been documented to vary significantly based on age and health of an individual (CDC, 2020). In this paper, we make use of this inherent variation to design a more effective set of restrictions that strategically releases some sub-groups of the population earlier than others.

The motivating question behind this paper is: How can we best protect the most vulnerable members of the population? Early death rates for COVID-19 have been alarming, particularly among the elderly. Infected individuals over the age of 65 face a fatality rate on the order of 10\% (Elflein, 2020). The first order concern here will be to reduce this death rate so that the elderly population is not decimated by this disease. For context, young individuals under the age of 35 face a barely measurable fatality rate of under 0.1\%, while the middle-aged cohort between ages 35 and 65 faces a fatality rate around 1\% (Elflein, 2020). 

At the time of the writing of this manuscript, a few cases of a new severe syndrome has been documented among children, possibly related to COVID-19 (Fleisler, 2020). While these cases are extremely rare so far, this occurrence highlights the potential uncertainty involved in evaluating the consequences of a plan for how to relax restrictions moving forward. It is difficult to ever state with certainty that we fully understand the risks involved in any plan, particularly one which involves releasing some segments of the population before others.

On the other hand, keeping the entire population under strict restrictions for an extended period of time has deleterious effects on the economy. Since restrictions were put in place in America, visits to commercial venues are down two-thirds (Couture et al., 2020), and this decreased activity has led to small business closure and layoff rates of around 50\% in Mid-Atlantic states (Bartik et al., 2020). Shutdown sectors represent over 20\% of all US payroll employment (Vavra, 2020) and the burden of these job losses has fallen primarily on the poor (Mongey et al., 2020).

Firms expect sales to fall by 6.5\% and uncertainty about future performance has risen over 40\% (Bartik, 2020), which has led to unprecedented stock market volatility (Baker, 2020). All told, the economic cost of closing non-essential businesses could total nearly \$10,000 per household per quarter (Mulligan, 2020), and this could cause persistent harm in the form of lower output and employment even after the shutdown ends (Huber, 2018).

While saving lives must be the primary concern, judiciously increasing economic activity earlier than otherwise is an important secondary goal as well. Perhaps the most important takeaway from the analysis in this paper is how to design policies that can responsibly mitigate the economic effects of a sustained shutdown without jeopardizing the most vulnerable members of the population.

We will make use of the inherent age-specific variation in the ability of hosts to handle infection to design disease control policies that minimize the number of severe cases that lead to deaths while increasing the ability of individuals to participate in activities that reduce the economic paralysis generated by this pandemic. This paper analyzes and explores the benefits of policies that are predicated upon around the timely release of younger individuals from social distancing restrictions before the release of older sub-populations. Compared to standard simultaneous release policies, carefully planned and executed staggered release polices are likely to lead to: 

\begin{itemize}
    \item Fewer infections over the entire duration of the pandemic
    \item Significantly lower death rate among the elderly
    \item Lower death rate across the whole population
    \item Increased economic activity at an earlier date without increased health risks
\end{itemize}

Of course, there is no perfect policy that fixes everything without any cost, so it is not surprising that some trade-offs will exist at the heart of our proposed social distancing framework. Specifically, within a staggered release policy, the group released first necessarily faces a higher immediate risk of infection (and by extension, death) than they would have if they had not been released.

To be concrete, the proposed release of the youngest group of individuals while older groups remain under strict restrictions might mean that expected infection and death rates among the youngest group would increase in the short run, relative to those that would result from the delayed simultaneous release of every individual.

On the other hand, the smaller population size at first release would also reduce overall risk to those released, so these trade-offs are complicated and bear further scrutiny. In some cases, the net effect of early release can actually reduce the risk faced by the young group overall, but this is not guaranteed. There are obviously important ethical considerations if first release were to expose the released population to higher risk and hence early return should not be mandatory.

However, it is important to remember that the recorded death rate of young individuals from COVID-19 thus far is on the order of 0.01\% (Elflein, 2020), and even the most severe staggered release policies considered in this paper would increase the estimated death rate for young individuals by less than 0.0001\%. For context, the same policies can decrease death rates for the elderly by over 1\%. 

To understand why staggered release policies would have such a strong impact on decreasing elderly death rates (and can even decrease young death rates), it is important to consider the exponential nature of any disease outbreak. Uncontrolled infections cause more and more infections, which leads to the disaster scenario of facing a large spike in infection rates that exceeds health care capacity. The desire to avoid a dangerous initial peak is precisely what informed the decision to install restrictions in the first place, and this motivation is eloquently captured in the familiar `flattening the peak' narrative (CDC, 2020).

However, the unique danger of COVID-19 lies precisely in its novelty. Severe restrictions, even if they are in place for an extended period time, will not prevent a dangerous second peak of infection rates because there was no chance for the population to build up any meaningful level of immunity before activity resumed (Feng et al., 2020). This is the downside of the any simultaneous release policy that serves as the benchmark for comparison in this paper. After the entire population is released, the infection once again may spread like wildfire. In particular, the resulting exposure of the elderly to high infection rates would lead to a higher density of severe cases that require intensive care, which threatens health care capacity and ensures greater loss of life.

The benefit of staggered release policies is exactly avoiding these dangerous levels of severe infections by redistributing both the likelihood of infection from the elderly to the young and from a more dangerous concurrent timing to a more manageable spread out timing. By releasing the young group from restrictions at an earlier date than others, their subsequent exposure to the disease, which would mostly result in asymptomatic or mild infections, provides a much higher density of immunity in the population when it comes time to release the older groups from restrictions. This lowers the likelihood of infection for everyone post-lockdown which should help avoid exceeding health care capacity. The concept of deploying recovered individuals with antibodies into society to restart the economy while keeping infectious contacts low is further developed in (Weitz et al., 2020) who term it `shield immunity.'

Compare this manageable level of infections to the potentially dangerous secondary peak that would appear under the simultaneous release policy. In that scenario, an overburdened health care system has significant negative spillover effects to all members of the population regardless of their COVID-19 infection status. Any patients with a severe infection during a time when health care systems are overburdened face a significantly higher risk of death, and it is precisely these additional deaths that staggered release policies are able to reduce, according to our results.

However, staggered release policies still need appropriate timing to achieve these benefits. In fact, the mistimed release of certain groups may lead to an increase in death rates, so this is not a trivial consideration. While the early release of the youngest group is generally going to be significantly helpful, the response of the epidemic curve to specific choices of when to release older groups is more complicated and requires more detailed analysis, as pointed out more generally by Morris et al. (2020).

As an example, suppose young individuals have already been released. The next decision may be when to release middle-aged individuals. Our results suggest that if this choice is timed to be just before the peak of infections generated by young individuals, this release actually significantly exacerbates the growth of infections to a new higher peak that can lead to noticeably higher death rates. By waiting until after the peak of infections generated by young individuals, this release only produces a small bump in infection rates that does not result in greater risks for anyone involved. This result is consistent with the findings in Morris et al. (2020).

As Morris et al. show, optimal control in this context responds to the changing levels of susceptibility and infection rate in the population. Our results show that, by waiting for the population to get past prescribed immunity levels, it is possible to ensure that each step in the staggered release policy has the intended positive consequences.

This paper is organized as follows: Section \ref{sec:Model} introduces the extended SEIR model. Detailed analysis and numerical simulations of the model are presented in Section \ref{sec:Results}, and the comparison of outcomes between staggered release policies and the simultaneous release benchmark is featured in this section as well. Section \ref{sec:Discussion} provides a discussion of the main results and concludes. Sensitivity analysis on the more uncertain parameters are included in the Appendix.

\section{Model}\label{sec:Model}

Simple SIR and SEIR types of epidemic models have been used in most current studies of COVID-19 dynamics (see, for example,  Li et {\it al.} 2020; Morris et {\it al.}, 2020; Sanche et {\it al.}, 2020; Weitz et {\it al.}, 2020). The model introduced in this section is an extension of the standard Susceptible-Exposed-Infectious-Removed (SEIR) model modified by the incorporation of asymptomatic infections, disease-induced deaths, hospitalizations, and preferential mixing between different age groups. These extensions enable a flexible description of COVID-19 disease transmission dynamics within a model that accounts for age-defined groups. Since we are modeling a single outbreak, other considerations such as aging, migration, births, and unrelated deaths are ignored.

\begin{figure}[htb!]
\centering
\includegraphics[width=280pt]{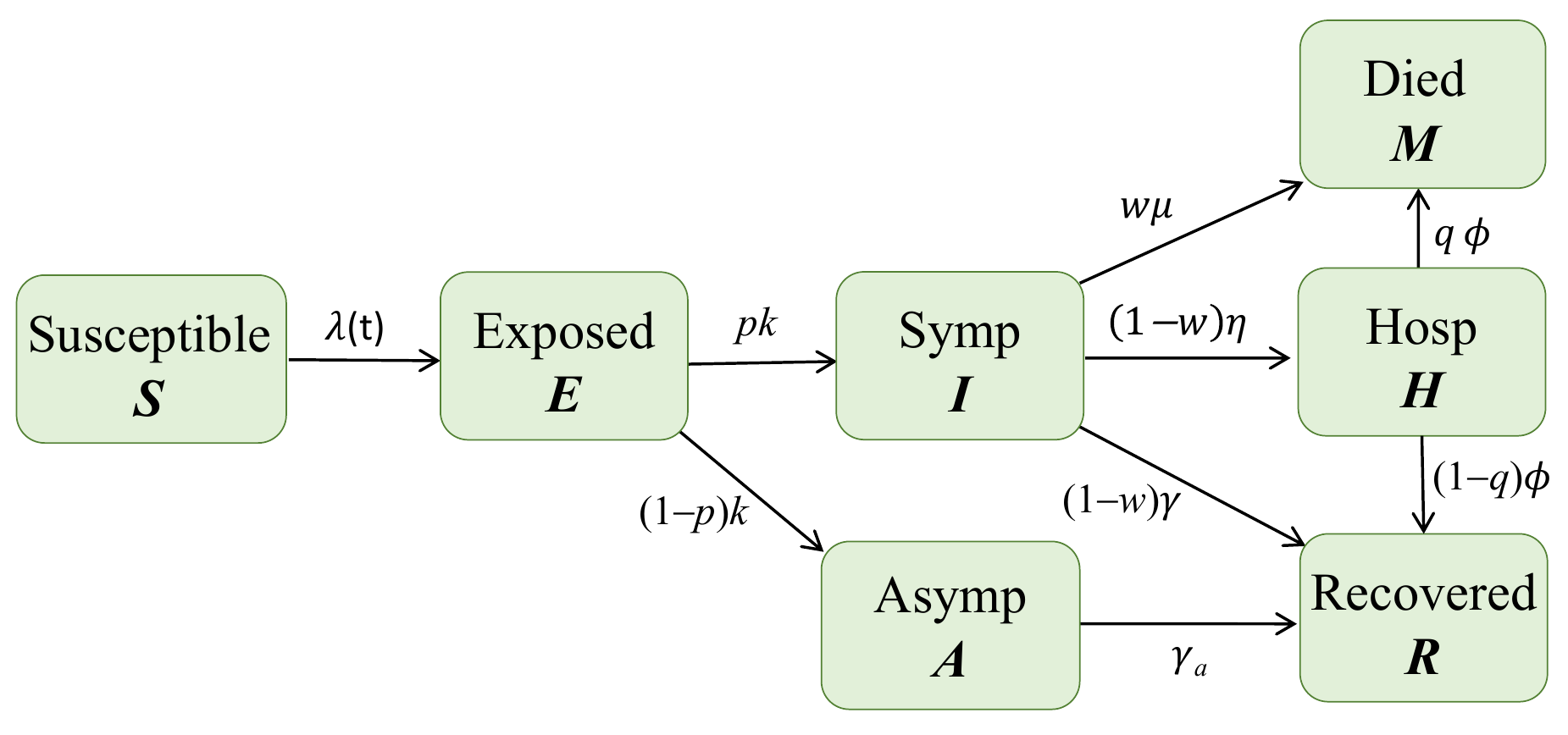}
\caption{
Depiction of disease transmission diagram corresponding
to Model \eqref{eq:model}. All sub-groups follow the same transition flows between
compartments, so the subscript $i$ is omitted. The infection rate $\lambda(t)$ includes
transmissions from infectious people in all sub-groups. The sub-groups are connected through a mixing function given in \eqref{eq:cij}.
 }\label{fig:diag}
\end{figure}

The population is divided into multiple age groups, also referred as sub-populations and labeled by  $i$ ($i=1,2, \cdots, n$). Each sub-population is further divided into seven epidemiological classes: susceptible $(S_i)$, exposed ($E_i$), infectious but asymptomatic ($A_i$), infectious and symptomatic ($I_i$), hospitalized ($H_i$), recovered ($R_i$), and dead due to disease ($M_i$). The total population is $N=\sum_{i=1}^n N_i$, where $N_i = S_i+E_i+A_i+I_i+H_i+R_i+M_i$ ($i=1,2, \cdots, n$). A disease transmission diagram for each sub-group is depicted in Figure \ref{fig:diag}.

The model consists of the following system of
differential equations:

\begin{equation}\label{eq:model}
\begin{array}{l}\vspace{.1in}
S_i'=- S_i \lambda_i(t), \\ \vspace{.1in}
E_i'=S_i \lambda_i(t) -k_i E_i, \\ \vspace{.1in}
A_i'=(1-p_i) k_i E_i-\gamma_{ai} A_i,  \\ \vspace{.1in}
I_i'=p_i k_i E_i-[(1-w)(\gamma_i+\eta_i)+w\mu] I_i, \\ \vspace{.1in}
H_i'=(1-w)\eta_i I_i-\phi_i H_i, \\ \vspace{.1in}
R_i'=\gamma_{ai} A_i+ (1-w)\gamma_i I_i+ (1_i-q_i(t)) \phi_i H_i, \\ \vspace{.1in}
M_i'=  q_i(t)\phi_i H_i+w\mu I_i, \ \ \ i=1,2, \cdots, n,
\end{array}
\end{equation}
where $\lambda_i(t)$ denotes the force of infection, the generator of new cases of infection among  susceptible individuals in group $i$. It is given by
\begin{equation}\label{eq:FOI}
\lambda_i(t)=\displaystyle  \sum_{i,j=1, i\neq j}^n a_i(t) c_{ij}
\frac{I_j + \theta_j A_j+\chi_j H_j}{N_j}, \ \ \ i=1,2,
\cdots, n.
\end{equation}
In  model (\ref{eq:model}), 
$k_i$ denotes the {\it per capita}
 rate of progression to the infectious state ($1/k_i$ represents the
 mean latent period, chosen to be 10 days),
 $\gamma_i$ and $\gamma_{ai}$
 denote the {\it per-capita} recovery rate if alive 
  with proportion $w$
 ($1/\gamma_i$ and $1/\gamma_{ai}$ are the
 mean infectious periods, chosen to be 7 days),
 $\eta_i$ denotes the rate at which symptomatic individuals
 are hospitalized if alive, $\phi_i$ is the rate at which hospitalized individuals leave the $H_i$ class with proportion $1-q_i$ 
 recovered and $q_i$ dead.
 $a_i(t)$ is  the {\it per capita} effective  contact rate, 
i.e.,
contacts that can lead to infection (Hethcote, 2000). In the case of no
intervention,  $a_i(t)=a_{0i}=\mathcal R_{0i} \gamma_i$, where
$\mathcal R_{0i}$ is the basic reproduction number for
group $i$. Our values for $\mathcal R_{0i}$ 
are based on the recent report in Sanche et {\it al.} (2020), which
gives an estimated value of the basic reproduction number to be
5.7 (95\% CI 3.8, 8.9). Although new estimates seem to suggest that these values are on the high range, we will be using them in our examples, since these values do not impact the overall objective of this paper, namely, that sequential age-dependent group releases offer us a framework to address the opening of the economy within a policy with the priority of protecting the most vulnerable.

Although our model is formulated for $n$ groups, in order to illustrate our approach, we consider three groups denoted by G1, G2, G3, representing young (age 35 and younger), middle-aged (35 to 65 years old), and elderly (65 and over), respectively.
It is assumed that the young group, G1, 
has a higher within-group basic reproduction number 
($\mathcal R_{01}$) than the older groups with themselves (G2 and G3).
We take the values  $\mathcal R_{01}=5$, $\mathcal R_{02}=3.6$, and
$\mathcal R_{03}=2.4$ in our numerical simulations.
When intervention measures are implemented, here via a
social distancing function that depends on the severity of the restriction, contact rates $a_i$ will be reduced.

The specific functional form of $a_i(t)$ used is given in \eqref{eq:a}.
For ease of reference, we will
refer to $a_i$ simply as a contact rate.
It is assumed in the model that among the
infectious people, proportions $p_i$ and $1-p_i$
are symptomatic and  asymptomatic, respectively. Asymptomatic 
and hospitalized individuals
can also transmit the disease but possibly at lower rates than symptomatic
individuals (references), which are denoted by the factors $\theta_i \leq  1$ and $\chi_i<1$, respectively. All parameters are non-negative and their definition is
also listed in
Table \ref{tab:1}. The numerical studies in this paper will focus on these 3 age groups.

The contacts between sub-groups are described
 by the mixing matrix $C=(c_{ij})$, where
 $c_{ij}$ is the proportion of the $i^{\hbox{\tiny th}}$ sub-group's contacts that is with members of the $j^{\hbox{\tiny th}}$ group.
 We will adopt the
 commonly used preferential mixing
(Nold, 1980 and later extended by 
Jacquez et {\it al.}, 1988), in which case, the elements of $C$ have the following form:
\begin{equation}\label{eq:cij} 
c_{ij}=\epsilon_i \delta_{ij}+(1-\epsilon_i) f_j,\quad \hbox{where} \ \ 
f_j=\displaystyle \frac{(1-\epsilon_j)a_jN_j}{\sum_k(1-\epsilon_k)a_kN_k},
\quad i, j=1, 2, \cdots, n,
\end{equation}
where
 $\epsilon_i \in [0,1]$ describe the preference level and $\delta_{ij}$ is  
the Kronecker delta (1 when $i = j$ and 0 otherwise).
The function $c_{ij}$ in \eqref{eq:cij} satisfies the required constraints 
for mixing functions (see Busenberg and Castillo-Chavez, 1991).

\begin{table}[htb!]
  \caption{Definition of the symbols used in
  Model \eqref{eq:model} and their values used in the simulations for the case of $n=3$ age groups..}
  \vspace{-.1in}
  \label{tab:1}
  {\small 
  \begin{tabular}{lll}
     \hline \smallskip
    Symbol & Description & Baseline Values\\
 \hline  
   $k_i$ & Rate of progression from $E_i$ to $I_i$, i.e.,   $1/k_i$ is the mean latent period  &
 1/10
 \\
   $\gamma_{ai}$ & Rate of recovery, i.e.,   $1/\gamma_{ai}$  is the mean infectious period for $A$ class  &
 1/7
 \\ 
 $\gamma_i$ & Rate of recovery, i.e.,   $1/\gamma_i$  is the mean infectious period for $I$ class  &
 1/7
 \\ 
 $\mathcal R_{0i}$ & Basic reproduction number of group $i$  &
$5, \ 3.6, \ 2.4$
  \\
 $a_{0i}$ & Effective contact rate in the absence of intervention  &
$\mathcal R_{0i} \gamma_i$
  \\
$a_i(t)$ & Effective contact rate under 
intervention such as social distancing &
See \eqref{eq:a}
  \\
$\theta_i$ & Infectivity ratio of asymptomatic
 to symptomatic individuals& 0.8
  \\
  $\chi_i$ & Infectivity ratio of asymptomatic
 to symptomatic individuals& 0.1
  \\
  $\epsilon_i$ & Level of preference for contacting one's own group, $0\leq \epsilon_i\leq 1$  &
0.7, \ 0.5, \ 0.9
  \\
  $c_{ij}$ & Proportion of contacts a member of group $i$ has 
   with
  group $j$& See \eqref{eq:cij}
  \\
  $\lambda_i(t)$ & Force of infection for susceptibles 
  in group $i$ at time $t$  & See \eqref{eq:FOI}
  \\
  $p_i$ & Proportions of infectious that are symptomatic. 
  $p_1\leq p_2 \leq p_3$&
0.25, \ 0.5, \ 0.8
  \\
 $q_i$ & Proportions of infectious that  
 die from the disease  &
See \eqref{eq:q}
  \\
  $s_b$ &  Reduction of contacts during the initial restrictions  &
0.8
  \\
  $s_i(t)$ & Reduction of contacts for group $i$ at time $t$  &
Vary
  \\
  $s_r$ &  Residual reduction of contacts after restrictions are lifted  &
0.1
  \\
   $T_1$--$T_4$ &  Time points when a new policy starts. $T_1<T_2<T_3<T_4$ & Vary
  \\
  $d_b$ & Duration of the initial period of restrictions  &
Varies  
  \\
  $d_s$ &   Duration for secondary period of restrictions  &
Varies
  \\
\hline
  \end{tabular}
  }\\
     Note: $i, j=1,2, \cdots, n$. Time unit is days. The three values are 
     in the order of groups 1,2,3.
\end{table}

Social distancing policies are introduced to reduce the standard contact rates.
Here, it is assumed that $a_i(t)$ is a function of time and  that their values
are influenced by the selected policy. Specifically, we model them via
 step  functions defined as follows:
\begin{equation}\label{eq:a}
  a_i(t) =\left\{ \begin{array}{ll}\vspace{.1in}
  \big(1-0.95s_b\big) a_{0i}, &T_1 \leq t < T_1+d_b, \\
\big(1-0.95 s_{i}(t)\big) a_{0i}, &  T_1+d_b < t < T_1+d_b+d_s, \vspace{.1in} \\
\big(1-0.95 s_{i}(t)\big) a_{0i}, &  T_1+d_b+d_s<t < T_1+d_b+d_s+d_r, \vspace{.1in} \\
\big(1-0.95 s_r\big) a_{0i}, &  t > T_1+d_b+d_s+d_r,
  \end{array}  \right.
  \end{equation}
where $T_1$ denotes the time when the initial restriction on everyone begin, $d_b$ represents the duration of the initial restriction, $s_b$ denotes the reduction of contacts during the initial restrictions from business as usual (0) to no contact (1), $d_s$ and $d_r$ represent the duration for secondary and tertiary restrictions which are characterized by $s_{i}(t)$ reductions of contacts for group $i$ for $t$ in that secondary and tertiary period, respectively, and finally $s_r$ represents residual restrictions after these policies are lifted.

The parameters $\mu_i$ and $\phi_i$ denote disease related death rates for
individuals in the $I_i$ and $H_i$ classes, respectively. To account for the dangers of overburdening health care systems, there will be a scaling factor added to the death rates of infected individuals as total infections approaches health care capacity. 
 The modified death rates $q_i(t)$ are thus functions of $H_i(t)$ with the following form:
\begin{equation}\label{eq:q}
  q_i(t) = q_{i0} + \sigma_i(\sum_{i=1}^3H_i(t)),
  \end{equation}
where $q_{i0}$ denotes the baseline level of 
disease mortality, and $\sigma_i>0$ is some convex function of the hospitalized infected population that captures the extra death rate caused by an increasingly overburdened health care system.

We will explore the effects of different policies by simulating the model with various schedules of social distancing restrictions prescribed by chosen paths for $s_i(t)$. Since the first month of the outbreak has already passed, it will generally be assumed that all policies considered include $s_b=0.8$, which corresponds to real policies used by governments around the world.

Finally, to allow for a direct comparison between outcomes, we will use projected death rates (likelihood of death) as the loss function. Note that $D_i=\int_0^\infty [\mu I_i(t)+q_i\phi_i H_i(t)]dt$ gives the total number of deaths for group $i$ over the course of the outbreak. Then let $\omega_i=\frac{D_i}{N_i}$ represent the proportion of group $i$ who dies, which we will refer to as the death rate for group $i$. Similarly defined, $\omega=\frac{1}{N}\sum_{i=1}^3 N_i\omega_i=\frac{1}{N}\sum_{i=1}^3 D_i$ is the overall death rate.

\section{Analysis and Results}\label{sec:Results}

This section describes the main results of this study for three age groups ($n=3$) and provides illustrations of the approach through simulations. Since there are multiple dimensions in the policy space, we first focus our attention on a set of polices with overall fixed duration. Recall that $d_b$ is the duration of initial restrictions and $d_s$ is the duration of secondary restrictions. We will consider policies with $d_b+d_s=d_T$ and, where $d_T$ is a fixed time where the secondary restrictions are assumed to end.

Further, since over a month\footnote{Simulations in this paper will assume the initial restrictions change after 30 days, but the results in this paper are robust to how long initial restrictions have been in place. Simulations using 90 days for this value are presented in the appendix in order to demonstrate that our results are not sensitive to this choice.} has already passed under restrictions, we take $d_b \geq 30$ with $s_b=0.8$ and since some residual restrictions may be required the value $s_r=0.1$ is assumed as well. The primary policy variable will be $s_i(t)$, the severity of restrictions on each group $i$ across time. In order to make the notation more accessible,  we  use $s_y(t) \equiv s_1(t)$ to denote the severity of restrictions on the young group (G1).  Similarly $s_m(t) \equiv s_2(t)$ and $s_o(t) \equiv s_3(t)$ for the middle-aged group (G2) and old group (G3).

While our notation allows for any progression of severity measures for $s_i(t)$, we will focus on particular policies that set a certain level of restrictions on each group across chunks of time, which we call the initial phase (which lasts for $d_b$ days), the secondary phase (which lasts for $d_s$ days), and the tertiary phase (which lasts for $d_t$ days). This structure was chosen because it allows for a fully staggered release of the population from restrictions one group at a time. In general, the idea will be to consider releasing the young group during the secondary phase and the middle-aged group during the tertiary phase.

\subsection{Early Release of the Young Group}\label{sec:YoungOnly}

First, we will consider policies that release the young group from restrictions before  other groups. There is no need to consider a tertiary period in this case, so these policies are represented by $s_y(t) < s_b$ for $t>T_1+d_b$, while $s_m(t) = s_o(t) = s_b$ for $t<T_1+d_b+d_s$. For ease of notation, we let $s_y^s = s_y(t)$ for all $ t \in [T_1+d_b, T_1+d_b+d_s]$ and refer to them as the severity of restrictions on the young group during the secondary period of restrictions. The range of policies available for consideration is $s_y^s \in [s_r, s_b] = [0.1, 0.8]$.

This class of policies includes all staggered release policies and $s_y^s=s_b=0.8$ represents the baseline simultaneous release policy, where all three groups are held under restrictions for a longer time before being released together. The goal will be to minimize death rates: $\omega_y$, $\omega_m$, $\omega_o$ and $\omega$. The following result shows how these death rates ($\omega_y, \omega_m, \omega_o, \omega$) respond to the early release of the young group from restrictions ($s_y^s<s_b$) by comparing outcomes from projected outbreaks under policies with various values of $s_y^s$.

\begin{result}
Overall death rate $\omega$ and death rates for the middle-aged and elderly groups, $\omega_m$ and $\omega_o$, are strictly increasing in severity of secondary restrictions $s_y^s$, while death rate for the young group $\omega_y$ is minimized at an intermediate value of severity $s_y^s < s_b$.
\end{result}

By fixing other aspects of the social distancing policy, such as the timing of intervention and severity of restrictions on other groups, we are able to identify the precise impact of releasing the young group from restrictions before other groups. The fundamental takeaway from this result is that releasing the least vulnerable group of a population before other groups will reduce the number of deaths overall.

\begin{figure}[htb!]
\centering
\includegraphics[width=265pt]{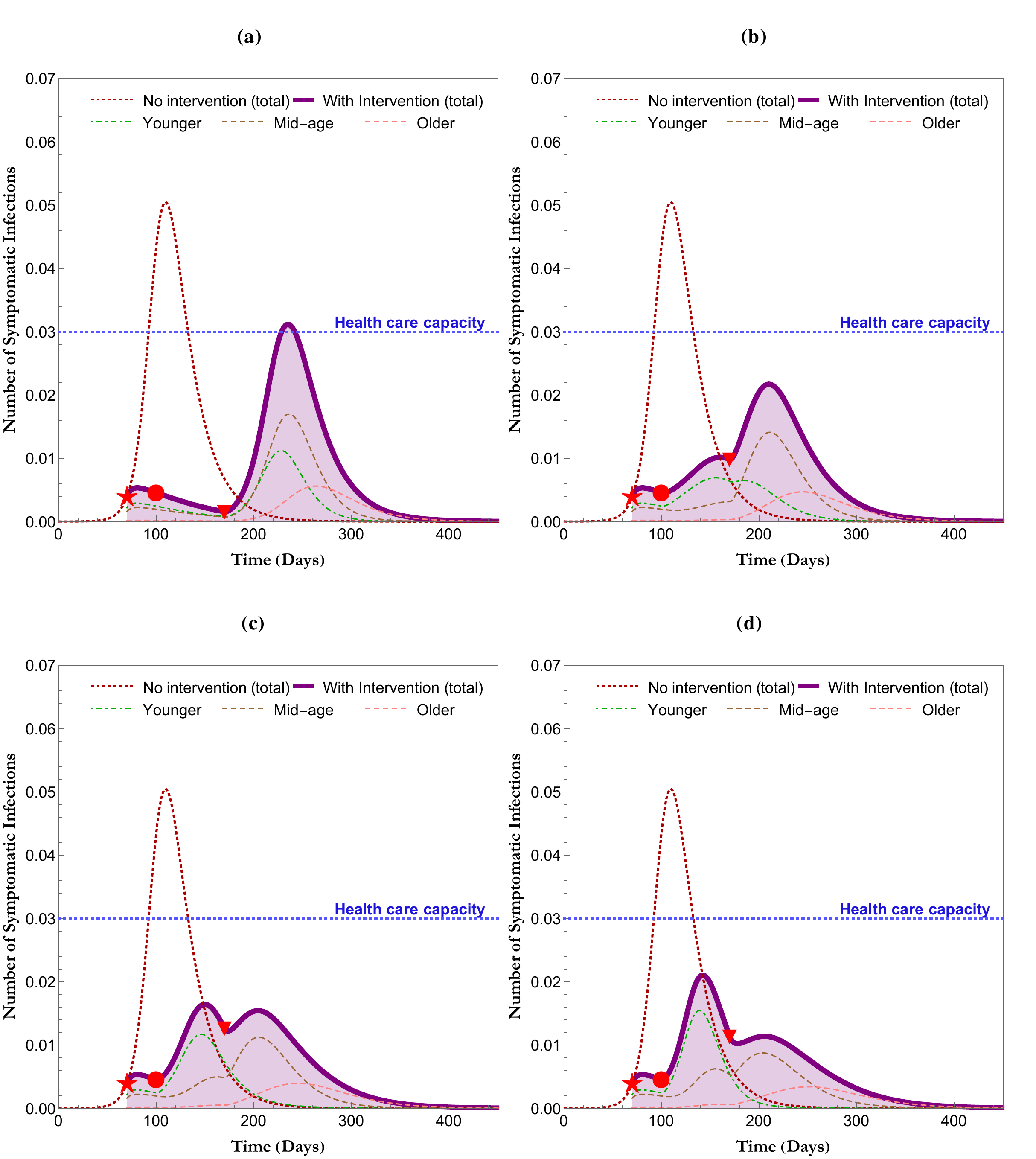}
\caption{
Comparison of projected outbreaks under the simultaneous release benchmark and three staggered release policies. Plot (a) shows the simultaneous release benchmark, with all three groups being released from restrictions at time $t=170$.
Plot (a) is the baseline scenario with 
$s_{y}(t)= 0.8$ for $t \in (T_1+d_b, T_1+d_b+d_s)=(100,170)$, which correspond to
the policy of extend the initial restriction of SD for an additional 70 days.
Plots (b), (c), and (d) show the cases when the young group has a relaxed policy
after the initial strict policy. More specifically, for $t\in (100,170)$,
$s_{y}(t)= 0.5, 0.3, 0.1$ in  (b)--(d), respectively. Other parameter values are given in the text.
 }\label{fig:scenarios}
\end{figure}

This strictly monotonic relationship between severity of secondary restrictions on the young and death rates for every group besides the young holds for all parameter values used in our simulations. A comparison between these policies is presented in Figure \ref{fig:scenarios} using one particular choice for these parameters: $T_1=70$, $d_b=30$, $d_s=70$, $s_b=0.8$, $s_r=0.1$, and $s_m(t)=s_o(t)=s_b$ for all $t < T_1+d_b+d_s$. That is, initial restrictions of severity $s_b=0.8$ were installed at time $T_1=70$ for a duration of $d_b=30$ days. Then at $t=T_1+d_b=100$, there is a choice of whether (and to what degree) to release the young group from restrictions; a choice of $s_y^s$. Regardless of this choice, the middle-aged and elderly groups are held under restrictions for $d_s=70$ more days, and eventually everyone is (mostly) released ($s_r=0.1$) from restrictions at $t=T_1+d_b+d_s=170$ days.

Figure \ref{fig:scenarios}(a) presents the baseline simultaneous release policy, which we use as a benchmark for evaluating efficacy of staggered release policies. It is the case where all groups are held under the same severity of restrictions for the secondary period and then all released at once at $t=170$ days. We observe that while infection rates are consistently decreasing throughout the 100 days when restrictions are in place, the lack of population immunity at the timing of release due to such strict restrictions leads to a dangerously high second peak of infection rates. This also carries significant risk of death for the most vulnerable population, the elderly, since they will need intensive care at a time when health care systems are overwhelmed.

Compare this benchmark simultaneous release policy with the other projected outbreaks under various degrees of staggered released policies. Figure \ref{fig:scenarios}(b-d), corresponding to $s_y^s= 0.5, 0.3, 0.1$ respectively, demonstrate the effect of staggered release: shifting the delayed peak earlier in time and decreasing the loss of life associated with that second wave. As the young group is released from restrictions, their activity leads to more infections earlier which helps in two ways.

The increased population immunity when the other groups are released lowers the risk of infection for more vulnerable groups, and if vulnerable members of the population do get sick at the later dates, there are fewer concurrent infections at that time so they can receive the level of care they need. By making infections both less likely and less dangerous for more vulnerable segments of the population, staggered release policies are able to significantly reduce death rates.

However, it may be important to note that this benefit is not without some small degree of cost. The increased infection rates incurred by members of the young group would mechanically lead to a slightly higher death rate under the most extreme staggered release policies ($s_y^s \leq 0.2$). While this is certainly a non-trivial trade-off, the actual impact is very small in absolute terms, on the order of 0.001\% higher death rates for $s_y^s=0.1$ compared to $s_y^s=0.8$.

\begin{figure}[htb!]
\centering
\includegraphics[width=340pt]{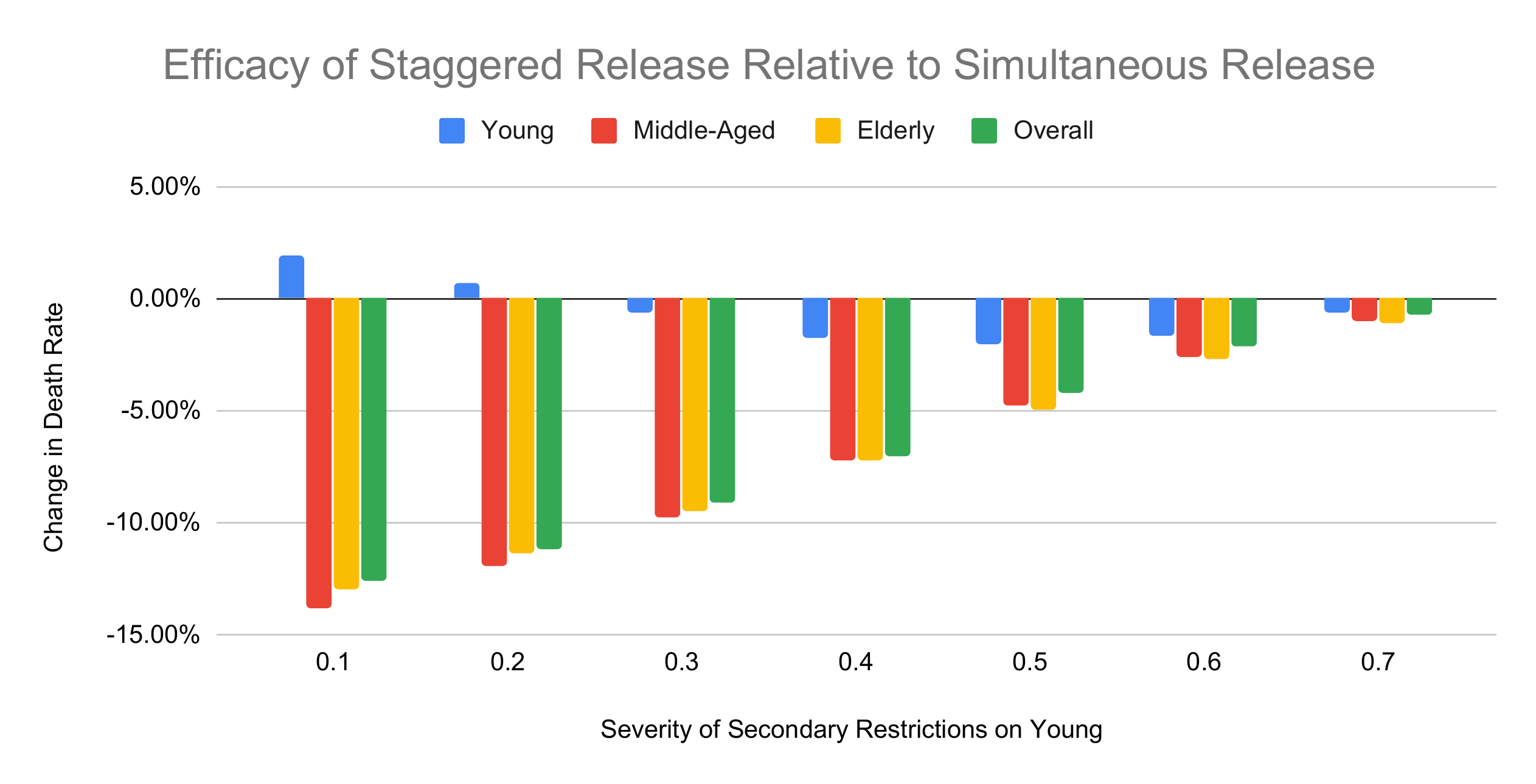}
\caption{Comparison of death rates in staggered release policies with the simultaneous release benchmark, measured as the percentage change in likelihood of death. This illustrates that compared to the simultaneous release benchmark, staggered release policies always decrease death rates for older groups and overall while sometimes decreasing death rates for the young group.
 }\label{fig:comp}
\end{figure}

Figure \ref{fig:comp} illustrates the behavior of death rates $\omega_y, \omega_m$, $\omega_o$, and $\omega$ in response to the range of choices $s_y^s < s_b$. Note the straightforward behavior of $\omega_m$, $\omega_o$, and $\omega$. These are all strictly decreasing as the young group is subject to less severe secondary restrictions. The change in $\omega_y$ is non-trivial, benefiting from some relaxed secondary restrictions but eventually this death rate increases as they face fewer secondary restrictions. Note that there are values of $s_y^s$ representing intermediate secondary restrictions that lead to decreased risk of death for all groups.

\subsection{Intermediate Release of the Middle-Aged Group}

The results in Section \ref{sec:YoungOnly} present the positive impacts of releasing the young group. In this section, we explore the impact of releasing the middle-aged group at some point between the young group and the elderly group as well. The intuition would be that since the middle-aged group is less vulnerable than the elderly group, staggering their releases may have similar benefits to the early release of the young group. We will observe a similar result here, but with some caveats.

In order to facilitate the intermediate release of the middle-aged group, we will now need to include a tertiary period of restrictions, during which that group may be released. This will be a natural extension of previous notation. As $d_b$ and $d_s$ referred to the duration of initial and secondary restrictions, $d_t$ will be used to refer to the duration of tertiary restrictions. As before, $s_o(t)=0.8$  for all $ t<T_1+d_b+d_s+d_t$, but now $s_m(t)=s_m^t \leq 0.8$ is another policy variable.

At this point it becomes expedient to denote four particularly important values for $t$. First, $T_1$ is the day of the outbreak on which initial restrictions are installed. We will maintain $T_1=70$ as before. Next, $T_2 =T_1+d_b$ is the day of the outbreak on which initial restrictions are lifted and secondary restrictions take place - for the policies in this section, the young group will be released at this time. Next, $T_3=T_2+d_s$ is the day of the outbreak on which secondary restrictions are lifted and tertiary restrictions take place - the middle-aged group may be released at this time. Finally, $T_4 = T_3+d_t$ is the day of the outbreak on which tertiary restrictions are lifted and every group is released at this time if they were not already.

More detailed analysis of the degree of release has already been done in section \ref{sec:YoungOnly}, so here we will focus on policies that either `release' a group ($s_i(t)=0.1$) or `do not release' a group from restrictions ($s_i(t)=0.8$) at each cutoff time $t \in \{T_2, T_3, T_4\}$. Thus policies can be further simplified as a choice of release dates ($r_y, r_m, r_o$) for the three groups, where $r_i \in \{T_2, T_3, T_4\}$. Note that the extremal staggered policy considered in section \ref{sec:YoungOnly} with $s_y^s=0.1$ would be represented by $(r_y, r_m, r_o) = (T_2, T_4, T_4)$ with $(d_b, d_s, d_t) = (30, 70, 0)$. We call  this the ``young release'' policy.

\begin{figure}[htb!]
\centering
\includegraphics[width=370pt]{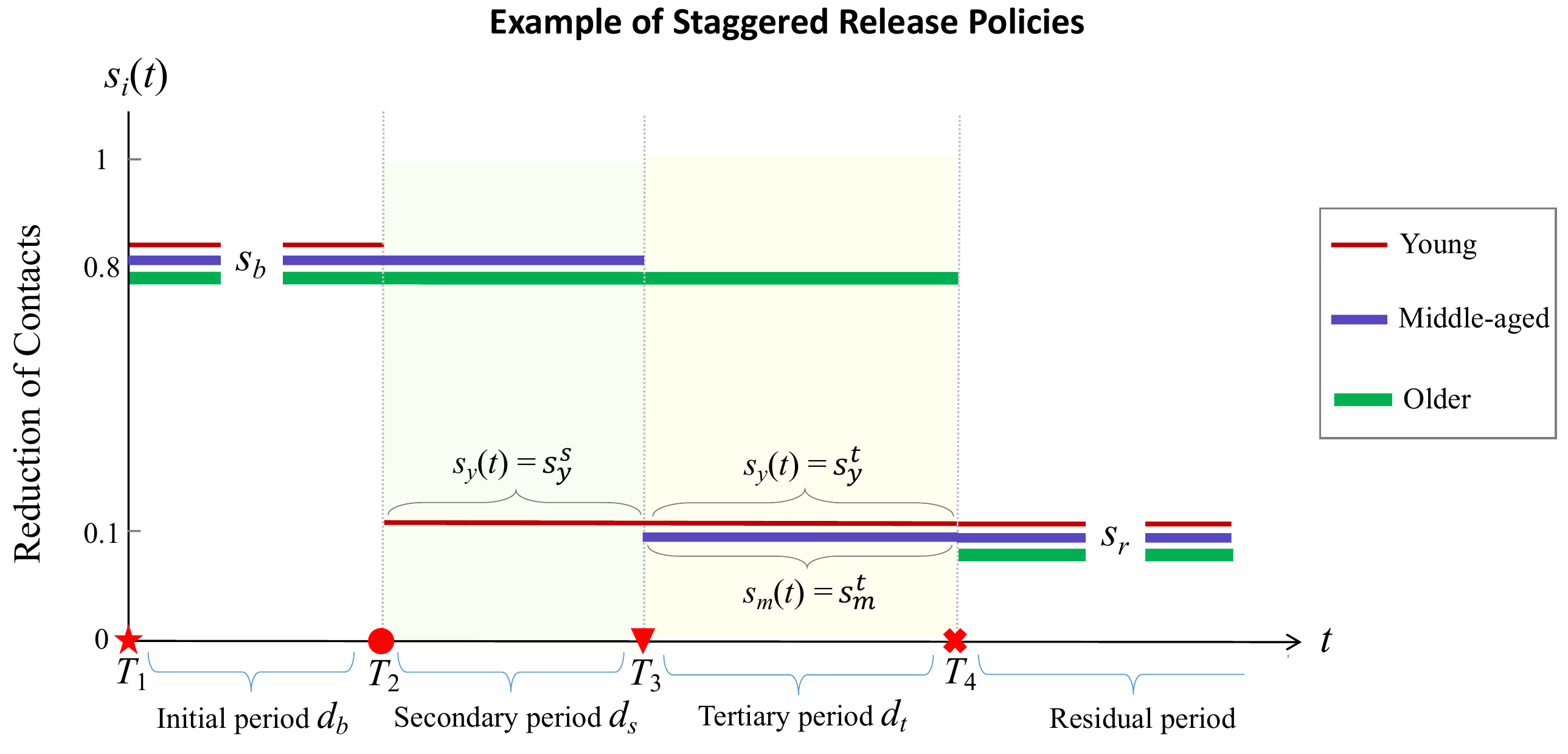}
\caption{
Depiction of one example showing the policy switching for the three age groups at times
$T_2, T_3,$ and $T_4$. This example corresponds to the scenario in which  the timing for policy switching are at
$(r_y, r_m, r_o)=(T_2, T_3, T_4)$ with levels of contact reductions: $s_b=0.8$,
$s_y^s=s_m^t=0.1$, and
$s_r=0.1$. This describes the staggered release strategies
illustrated in Figures \ref{fig:comp3G}(c) and (d).
 }\label{fig:example}
\end{figure}

The baseline simultaneous release policy will be the same idea as before, with all three groups being held under the same severity of restrictions as the initial restrictions until $T_4$. This is represented by $(r_y, r_m, r_o) = (T_4, T_4, T_4)$. Two natural comparisons are the ``young release'' policy and the ``middle release'' policy of $(r_y, r_m, r_o) = (T_2, T_3, T_4)$, where the young group is released after initial restrictions and the middle-aged group is released after secondary restrictions. Figure \ref{fig:example} depicts the timing of policy changes for an example staggered release policy. These policies will be evaluated based on the resulting peak size of infections and projected death rates for the elderly ($\omega_o$) and overall ($\omega$).

\begin{figure}[htb!]
\centering
\includegraphics[width=300pt]{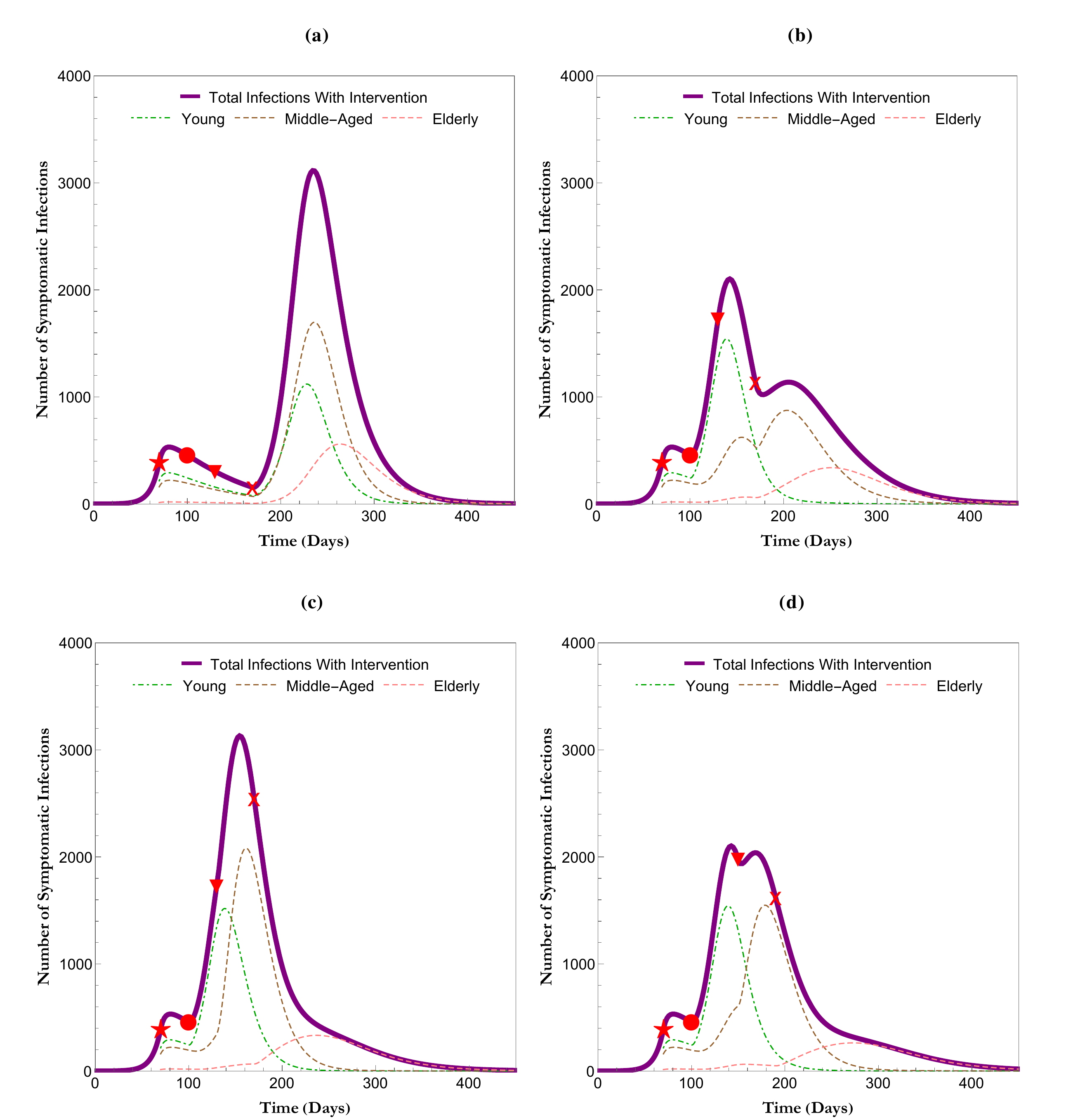}
\caption{Total symptomatic infections under four different policies. Plot (a) is the baseline case where no groups are released until $T_4=170$ days. Plot (b) is the ``young release'' case where the young group is released at $T_2=100$ days. Plot (c) is the ``young and middle-aged release'' case, where the young group is released at $T_2=100$ days and the middle-aged group is released at $T_3=140$ days. Plot (d) is the ``alternate staggered release'' case, where the young group is released at $T_2=100$ days and the middle-aged group is released at $T_3=150$ days. More detailed descriptions of these results can be found in the text and Table \ref{tab:4policy}.
 }\label{fig:comp3G}
\end{figure}

Figure \ref{fig:comp3G} shows projections for these three policy choices with $(d_b, d_s, d_t) = (30, 30, 40)$ along with a fourth policy which is a slight variation on the ``middle release'' policy, with $(d_b, d_s, d_t) = (30, 50, 40)$. This fourth policy is an adapted version of the ``middle release'' policy designed as a minimal change in order to fix a key oversight that would lead to negative consequences if the ``middle release'' policy is implemented. These four policies along with projections for peak size and various death rates are summarized in Table \ref{tab:4policy}.

\begin{table}[htb!]
    \centering
    \begin{tabular}{c|c|c|c|c}
         & (a) & (b) & (c) & (d) \\
         \hline
        Timing ($d_b, d_s, d_t$) & (30, 30, 40) & (30, 30, 40) & (30, 30, 40) & (30, 50, 40) \\
        \hline
        Release ($r_1, r_2, r_3)$ & ($T_4, T_4, T_4$) & ($T_2, T_4, T_4$) & ($T_2, T_3, T_4$) & ($T_2, T_3, T_4$) \\
        \hline
        Peak Size (per 100,000) & 3116 & 2103 & 3132 & 2040 \\
        \hline
        Death Rate ($\omega_o, \omega$) & (6.62, 1.43) & (5.76, 1.25) & (5.62, 1.27) & (5.25, 1.19) 
    \end{tabular}
    \caption{Comparison of four different policies and associated outcomes. A detailed description of these policies can be found in the text and Figure \ref{fig:comp3G} offers a visualization of their projections.}
    \label{tab:4policy}
\end{table}

As before, the change from the baseline case (Figure \ref{fig:comp3G}a) to the ``young release'' case (Figure \ref{fig:comp3G}b) is significantly positive in terms of reducing peak size and reducing death rates for the elderly and overall. However, the ``middle release'' case (Figure \ref{fig:comp3G}c) does not correspond to an unambiguously positive change when compared to the impact of releasing  the young' case. This second release  reduces the death rate among the elderly albeit it results in a higher death rate overall. It leads to a higher peak size than even the simultaneous release benchmark.

The reason for this detrimental effect on overall death rate can be seen from Figure \ref{fig:comp3G}. Since $T_3$ occurs close to a peak in the infection rate, releasing the middle-aged group at that point puts them at significant risk of infection and death, even though they are much less vulnerable in general than the elderly. This particular timing actually exacerbates the peak size of the outbreak endangers the middle-aged group more than it protects the elderly. While the projected death rates in this case are still lower than in the simultaneous release benchmark, the peak size is actually higher.

Fortunately, these downsides can be avoided by waiting some additional time before releasing the middle-aged group, which we will call the ``patient staggered release'' policy (Figure \ref{fig:comp3G}d). By waiting for infection rates to begin to decrease, the release of the middle-aged group causes only a minor spike in infections,  small enough that it does not  jeopardize members of the group. While this does result in a slightly higher risk for the middle-aged group since they were still released earlier than in the ``young release'' case, this fourth policy meaningfully reduces elderly and overall death rates, as well as decreasing peak size (see Table \ref{tab:4policy}). This relationship between timing of release for intermediate groups and the path of the outbreak informs our second main result:

\begin{result}
Assume the young group has already been released at $T_2$. There exists a cutoff $T^*$ such that releasing the middle-aged group at $T_3<T^*$ increases death rate $\omega$ and releasing the middle-aged group at $T_3>T^*$ decreases death rate $\omega$. This $T^*$ is closely related to the time at which infection rates begin to decrease.
\end{result}

\begin{figure}[htb!]
\centering
\includegraphics[width=400pt]{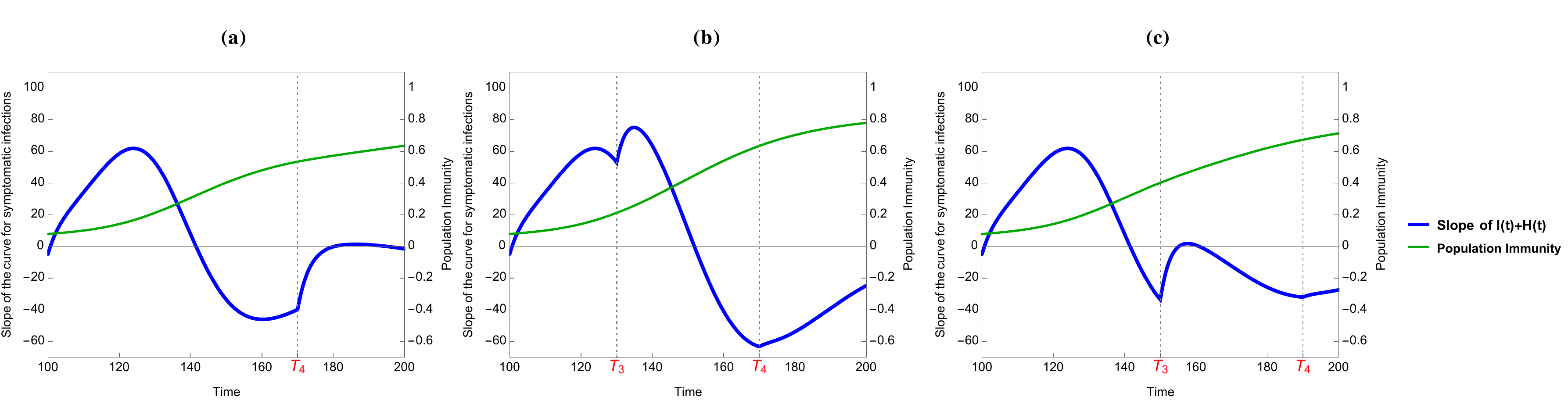}
\caption{Plots of the slope of the function, $I(t)+H(t)=\sum_{i=1}^3\big(I_i(t)+H_i(t)\big)$ (thick blue curve), and the population immunity (thinner green curve), 
$R(t)=\sum_{i=1}^3R_i(t)$. The figures in (a) and (b) used the same parameter values as in  Figure \ref{fig:comp3G}(c) and (d), respectively. It shows that
the slope at $T_3$ is positive in (b) and negative in (c), with the immunity levels around 17\% and 34\%, respectively. It also shows that the slope is zero at $t=142$ days, suggesting that the middle-aged group should be released after that time.
 }\label{fig:immunity}
\end{figure}

The importance of properly timing intermediate release of the middle-aged group can be explained by examining the $I_i$ and $H_i$ equations in Model \eqref{eq:model}. The total number of symptomatic infections at time $t$ is given by $I(t)+H(t)=\sum_{i=1}^3\big(I_i(t)+H_i(t)\big)$, and the nature of how this curve responds to a subsequent group's release is what determines whether that decision is beneficial or harmful overall. The slope of this function $I(t)+H(t)$ is presented in Figure \ref{fig:immunity}, along with the function for population immunity at time $t$, $R(t)=\sum_{i=1}^3R_i(t)$.

Figure \ref{fig:immunity}(a) shows the slope of the infection curve under the ``young release'' policy depicted in Figure \ref{fig:comp3G}(b), which we can observe begins positive from $T_2=100$ when young people are released but is eventually negative at some point between $T_2$ and $T_4$ (there is no meaningful $T_3$ here since middle-aged people are not released early in this policy). This point when the slope goes from positive to negative will be closely related to $T^*$, the cutoff point when releasing the middle-aged group becomes beneficial. 

Figure \ref{fig:immunity}(b) shows the slope of the infection curve under the ``middle release'' policy depicted in Figure \ref{fig:comp3G}(c). Here we can observe the choice of timing $T_3$ is too early; releasing middle-aged individuals at this value of $T_3$ exacerbates the infection rate and this effect is what leads to higher death rates overall. Compare this to Figure \ref{fig:immunity}(c), which shows the slope of the infection curve under the ``patient staggered release'' policy depicted in Figure \ref{fig:comp3G}(d). This choice of $T_3$ is after the slope of the infection curve has become negative, and it does not cause a large spike in infections.

Since it is generally difficult to accurately measure concurrent infections, one useful proxy is the population immunity rate. The green curves presented in Figure \ref{fig:immunity} show the recovered portion of the population. In order to identify the correct choice for $T_3$, it would be possible to test members of the population for antibodies in order to conclude enough population immunity exists to believe the next step of the staggered release policy would be beneficial. Note that the measured population immunity at the choice of $T_3$ in the ``middle release'' policy visible in Figure \ref{fig:immunity}(b) would be below 20\%, while the corresponding population immunity at $T_3$ in the ``patient staggered release'' policy would be closer to 40\%. This is an operational method to choose $T_3$ correctly.
 
\begin{figure}[htb!]
\centering
\includegraphics[width=330pt]{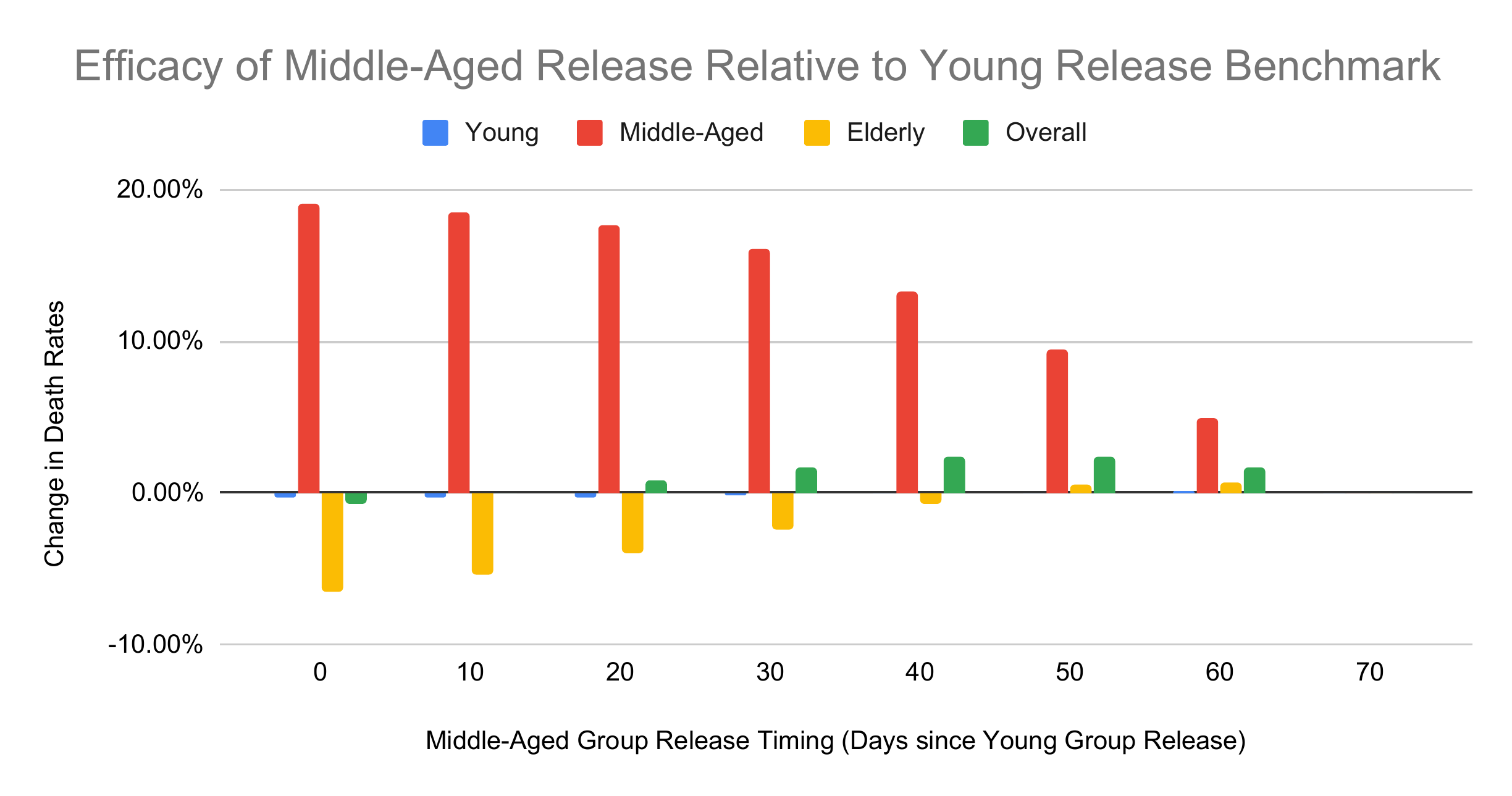}
\caption{Comparison of death rates when middle-aged group is released earlier than the elderly group, measured as the percentage change in likelihood of death. This illustrates that compared to the ``young release'' benchmark, releasing the middle-aged group may decrease the elderly group's death rate but sharply increases the middle-aged group's death rate.
 }\label{fig:comp2}
\end{figure}

However, there is still an important takeaway from the fact that the ``middle release'' policy had significantly negative consequences. This reveals the importance of properly considering the timing of any staggered release policy, since the benefits of fostering immunity through release of less vulnerable groups can be outweighed by the immediate exposure of more groups to high infection rates. Figure \ref{fig:comp2} shows how severe these consequences can be when staggered release is timed wrong.

Releasing the middle-aged group too soon after releasing the young group puts more people into the active population at a time when infection rates are high, and while the middle-aged group is much less vulnerable than the elderly group, they are also much more vulnerable than the young group so these increased infections lead to a significant increase in deaths. This leads to the discouraging result that the earlier this middle-aged group is released, the higher the death rate they face will be.

However, even in this particularly stark case, the reduction in death rates for the elderly can be strong enough to reduce death rates overall, since the elderly death rate sits at a much higher magnitude than the middle-aged death rate. This leads to an interesting implication that releasing middle-aged members of the population only really hurts themselves while marginally benefiting others, so perhaps a voluntary release policy could have some merit in this context. Such questions are likely outside the scope of this paper, however.

\begin{figure}[htb!]
\centering
\includegraphics[width=330pt]{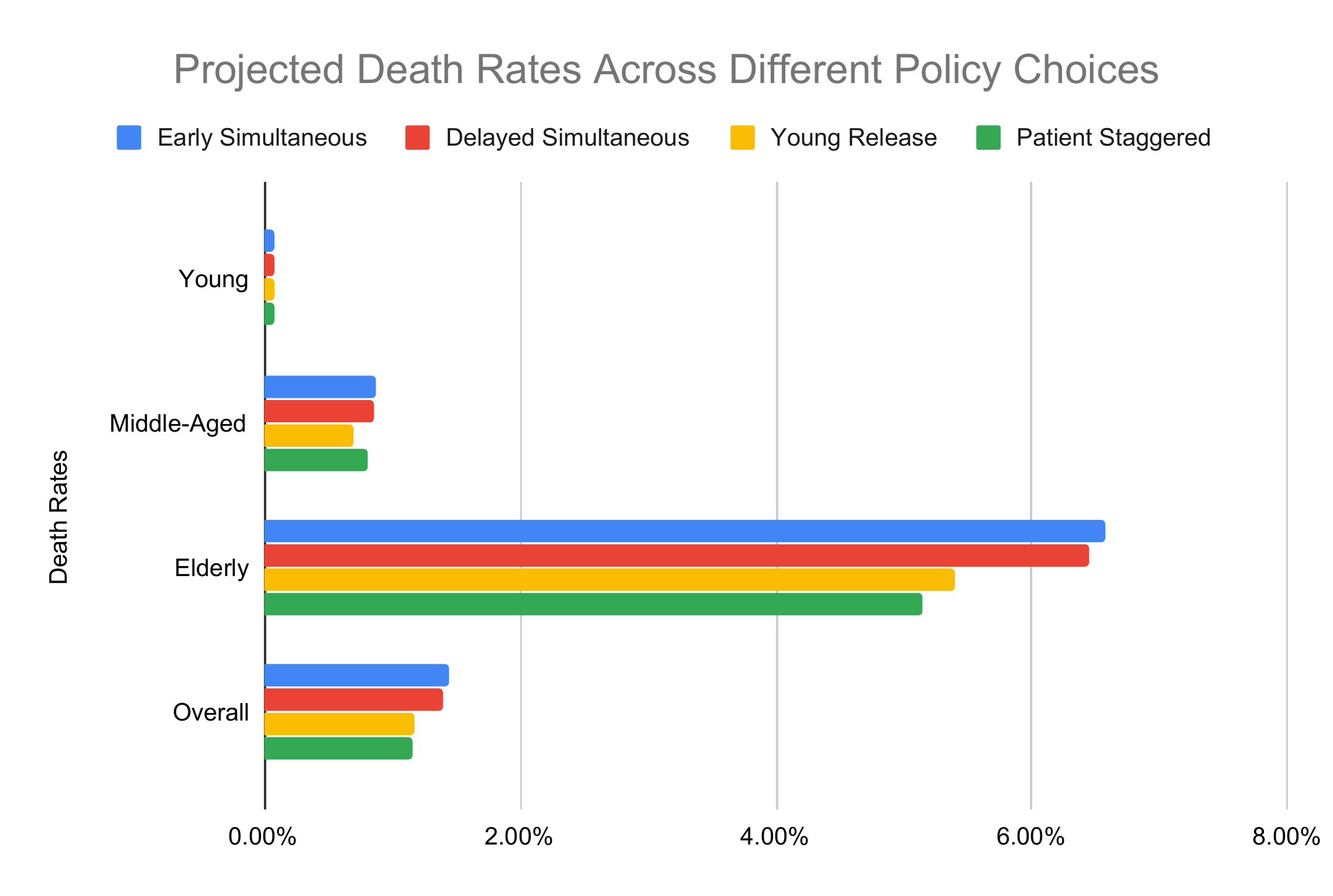}
\caption{Comparison of death rates projected under the ``Early Simultaneous Release'' policy (everyone released after thirty days of restrictions), the ``Delayed Simultaneous Release'' policy (everyone released after 110 days of restrictions), the ``Young Release'' policy (young released after thirty days of restrictions, middle-aged and elderly after 110), and the ``Alternated Staggered Release'' policy (young released after thirty days of restrictions, middle-aged after eighty, and elderly after 120). In general, the two staggered release policies lead to better outcomes than either simultaneous release policy. More discussion on this can be found in the text.
 }\label{fig:PolComp}
\end{figure}

Finally, another important benchmark may be the one generated by the types of policies being considered and enacted in some states, where the entire population are being  already released from restrictions. Under our framework here, this would be represented by release dates for all three groups being immediately after initial restrictions, so $(r_y, r_m, r_o) = (T_2, T_2, T_2)$. This carries significant risk of a dangerous second peak without any groups being protected, so it will result in much higher death rates for the elderly and overall. A comparison between death rates for this ``early release'' policy and the other policies considered in this section is presented in Figure \ref{fig:PolComp}.

First and foremost, the death rates for every single group is the highest under the early release policy. Perhaps interestingly, the delayed simultaneous release policy does not perform much better than the early release policy in terms of these projected death rates either. This relationship reinforces the point that an essential component to any successful disease control policy must consider some version of staggered release across groups of the population or risk simply delaying the inevitable. Since strict measures imposed on the entire population does not allow for any immunity to build up in the interim, there is no significant benefit to those restrictions.

However, the two staggered release policies perform much better than either simultaneous release option. The death rates for middle-aged people is minimized under the young release option, while the elderly death rate and overall death rate are minimized under the patient staggered release option. This represents a non-trivial trade-off between decreasing death rates for the middle-aged group and the elderly, but whichever choice is made would result in one of the staggered release policies being chosen. Since members of the young group face very small risk from the disease, only negligible changes to their death rate occurs across any of these policies.

The more fundamental question of how to handle the trade-off between economic stagnation and health risks is certainly difficult to answer, but from these results it would appear that however that problem is ultimately tackled, the resulting optimal policy must incorporate some form of staggered release. An early simultaneous release is far too costly in terms of health risks and loss of life to justify the marginal increase in economic activity compared to even a slightly staggered release policy. On the other hand, a delayed simultaneous release is both economically detrimental and more dangerous than staggered release policies.

\section{Discussion}\label{sec:Discussion}

The results in this paper support the possibly counter-intuitive notion that releasing some groups earlier from social distancing restrictions may in fact benefit everyone involved. While standard policies under consideration almost always treat the entire population as one group for the purposes of enacting or retracting restrictions, the optimal policy should actually treat sub-groups of the population differently and release them from restrictions sequentially in order to best protect vulnerable members of the population while also promoting economic recovery.

It should be noted that the age distribution used in the analysis within this paper is appropriate for the population of the USA, but applying the same framework to other countries may require significantly different population assumptions. For instance, countries like Italy, Japan, or Spain would have a higher proportion of elderly individuals. While the qualitative results should still hold, any quantitative conclusions would obviously be sensitive to the age structure of the population.

One of the main results is that releasing the youngest group of the population, which in the model considered here was the section of the population under the age of 35, will not only significantly reduce death rates among the elderly, but also reduce death rates for the middle-aged group and potentially for the young group as well. This type of staggered release policy succeeds in these goals by preventing a dangerously high second peak from occurring while protecting the most vulnerable group, the elderly, by allowing a higher degree of immunity to build up in the population before they are released.

Compared to the delayed simultaneous release benchmark, staggered release policies that release the young group first can save a significant portion of lives among the elderly. In the particular example presented in Figure \ref{fig:comp}, the magnitude of this effect exceeds a 10\% reduction in elderly death rates, which would translate to over 1\% of the elderly population being saved, since the elderly face a recorded 10\% death rate (Elflein, 2020). The estimated size of this effect is robust to alternate parameter specifications, and some sensitivity analysis is presented in the Appendix. For comparison, the corresponding estimated increase in the death rate of young individuals in this example is on the order of 0.0001\%.

While 0.0001\% is still positive, there are many aspects of society which involve some group accepting (in some cases, significantly) greater personal risk in order to reduce risks faced by others. Some examples include military service, first responders, and  health care workers during this pandemic. For young individuals under a staggered release policy, this slight increase in risk would also be compensated by the ability to return to work and everyday life at a much earlier date. 

Furthermore, even a staggered release policy that releases young people from restrictions would not require every single person under the age of 35 to resume activity. The already small death rate observed among the young is likely driven primarily by a subset who are immunocompromised or otherwise find themselves at higher risk, and these individuals can elect to remain at home. If these choices were taken into account, it is likely that even the most severe staggered release policies would not increase the death rate among the young.

Another important result demonstrates the importance of properly timing the release of subsequent cohorts in order to optimally control the progression of an outbreak. Comparing Figure \ref{fig:comp3G}(c) with Figure \ref{fig:comp3G}(d) reveals the stark change in outcomes if the release of middle-aged individuals is mistimed to be even a few weeks too early. Releasing these individuals from restrictions while infection rates are still rising results in a significantly higher peak size of infections and much higher death rates compared to waiting and releasing them after infection rates have begun to fall.

Importantly, while economic considerations were mentioned as a motivating factor behind the earlier removal of restrictions for some groups, there were no actual economic components in the model we used in this paper. Our conclusion that staggered release policies outperform simultaneous release policies is shown through a direct comparison of health risks and aggregate loss of life estimated from projections of the outbreak under these policies. The fact that earlier release of certain sub-groups enhances economic activity on the whole is a further benefit to staggered release policies, but this is a benefit that exists outside of our model.

We want to stress that we do not use increased economic activity to justify increasing health risks or loss of life. Staggered release policies accomplish both goals rather than putting either one before the other. We also want to stress that while the analysis in this paper provides support for the idea of releasing some people earlier, it absolutely does not provide support for any plan to end all social distancing policies prematurely.

The very idea of a staggered release policy is to only release part of the population, and releasing everyone early would lead to disastrous results. The analysis we perform relies on the continuation of restrictions on the older and more vulnerable group(s) in order to achieve better outcomes for them in particular. Releasing these groups early as well would result in a higher risk of death for everyone and would never be justified by the type of analysis done here.

Additionally, the underlying parameters used in this paper rely on current knowledge and beliefs about COVID-19 -- preferential mixing parameters, proportion of asymptomatic cases, infectious period, etc. The commonly accepted values for these may change in significant ways as we learn more about the disease, but our qualitative results hold across a wide variety of choices for these parameters. This sensitivity analysis is presented in the Appendix.

Ultimately, it was our goal to analyze the idea that staggered release policies may have some meaningful advantages over a delayed release for everyone. While there must certainly be more analysis with much greater precision before we could confidently recommend these policies be implemented, we do believe there is significant merit to their consideration. Even before economic concerns are inserted into the discussion, it appears that there can be direct benefits to all parties involved in terms of a lower risk of infection and death. When such concerns are incorporated, staggered release policies would thoroughly outperform simultaneous release policies.

\subsection{Conclusion}

Social distancing restrictions have been a powerful and important tool in controlling the spread of COVID-19 and keeping infection rates low enough to avoid exceeding health care capacity. Now that the initial phase of the outbreak has been addressed, the next step is to properly schedule the relaxation of constraints in order to optimally control the entirety of the outbreak. The decision to lift restrictions on everyone simultaneously is perhaps the natural policy, but this can have dire consequences if a second peak arises as a result of insufficient immunity in the population. Such an outcome would defeat the purpose of installing restrictions in the first place, so it is crucially important to consider more effective policies in terms of when to release people from restrictions.

To summarize, the early release of young individuals provides an interim period that allows them to build up some degree of immunity without facing significant health risks themselves, in order to protect older and more vulnerable individuals from severe health risks by reducing their likelihood of infection and death. As an added benefit, the staggered release policy allows for more activity in the population at an earlier date than the delayed simultaneous release policy would. Thus staggered release policies can positively impact everyone involved by allowing people to return to their lives without risking them.

\section*{Acknowledgments}
ZF's research is partially supported by NSF grant DMS-1814545.
CCC research is partially supported by NSF Expeditions in Computing Grant CCF-1918656,  CCF-1917819.

\section*{Disclaimer}
The findings and conclusions in this report are those of the authors and do not necessarily represent the official views of the National Science Foundation.

 \begin{appendix}
 
\section{Appendix}

We include in this appendix 3 examples to demonstrate that the main conclusions discussed in this paper are not sensitive to the choice of those 
model parameters that have high level of uncertainty. These include
the proportions of symptomatic infections ($p_i$), levels of preferential mixing ($\epsilon_i$), and the recovery rate $\gamma$. 
 We will present the aggregate death rates between a staggered release policy and a simultaneous release policy for a variety of values for each of the following parameter groups:  proportion of asymptomatic cases ($p_i$), 
 preferential mixing ($\epsilon_i$), recovery rate ($\gamma$), and the duration of restriction for the initial period ($d_b$).

\subsection{Proportion of asymptomatic infections, $p_i$}

Consider first the parameters $p_i$. The values used in the simulation presented in the main text are
 $(p_y, p_m, p_o) = (0.25, 0.5, 0.8)$.
To check how variation of these parameter values may affect the conclusions, such as the disease-induced death rates.
we examined the following sets: $(p_y, p_m, p_o) = (0.25, 0.25, 0.25), (0.5, 0.5, 0.5), (0.8, 0.8, 0.8), (0.35, 0.25, 0.45)$, and $(0.1, 0.2, 0.3)$. All other parameters are the same as in  Figure \ref{fig:PolComp}. Results are shown in Figure \ref{fig:sens}(a), and we observe qualitatively the same
behavior as in Figure \ref{fig:PolComp}.

\subsection{Preferential mixing parameters, $\epsilon_i$} 

The parameter values used in  the main text are $(\epsilon_y, \epsilon_m, \epsilon_o) = (0.7, 0.5, 0.9)$. Consider the following  choices for these values: $(\epsilon_y, \epsilon_m, \epsilon_o) = (0.5, 0.5, 0.5), (0.9, 0.9, 0.9), (0, 0, 0)$, and $(0.35, 0.25, 0.45)$.
 All other parameters are the same as in  Figure \ref{fig:PolComp}. This is presented in Figure \ref{fig:sens}(b).
Again,  we observe qualitatively the same
behavior as in Figure \ref{fig:PolComp}.

\subsection{Recovery rate $\gamma$ of symptomatic infections} 

\begin{figure}[htb!]
\centering
\includegraphics[width=280pt]{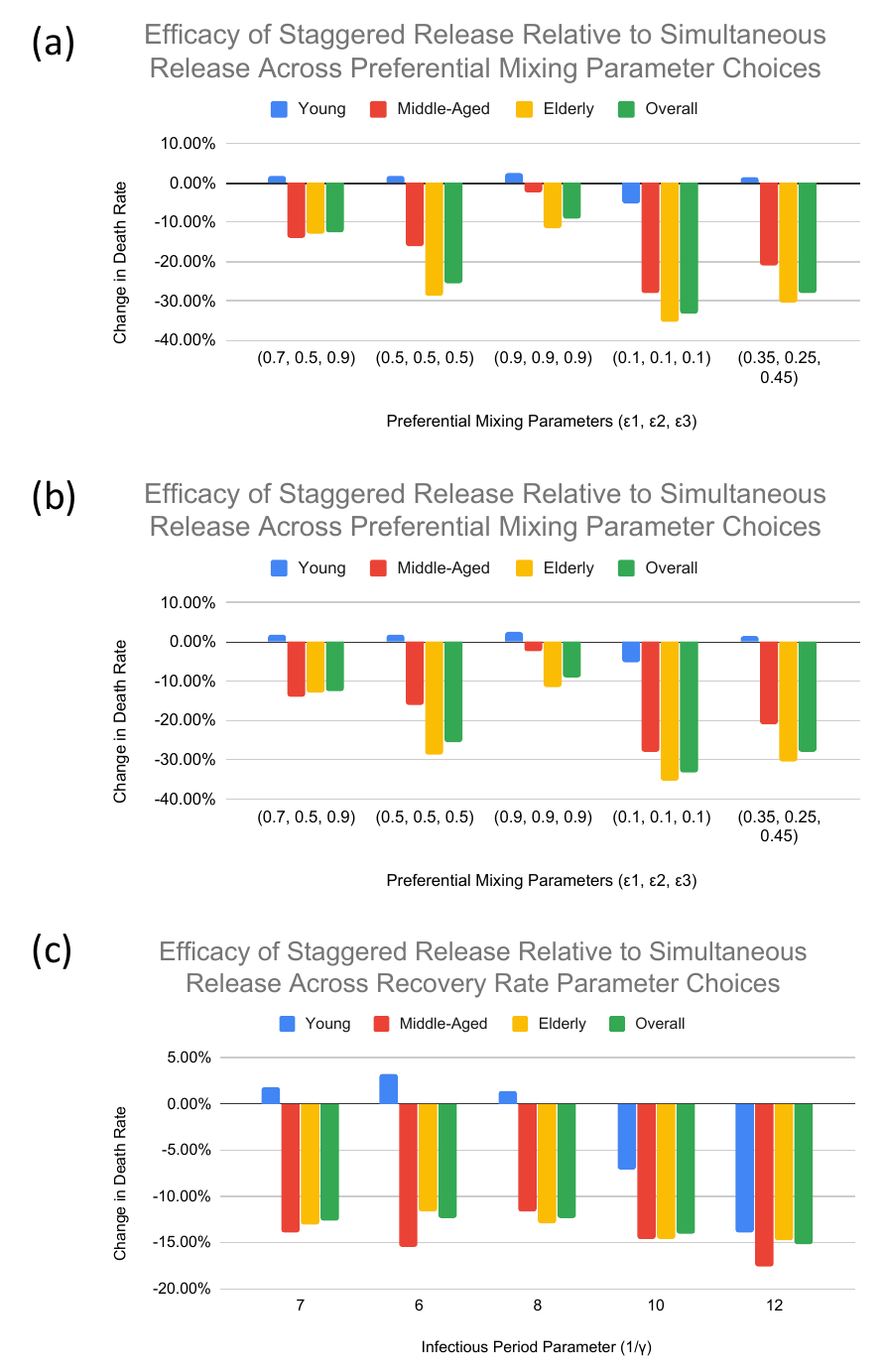}
\caption{
Sensitivity analysis for (a) Symptomatic Proportion ($p_i$),
(b) Preferential mixing parameters ($\epsilon_i$), and
(c) Recovery rate of symptomatic infections ($\gamma$).
 }\label{fig:sens}
\end{figure}

In the main text, we used $1/\gamma = 7$. For robustness, we will examine: $1/\gamma = 4, 6, 8,$ and $10$.  All other parameters are the same as in  Figure \ref{fig:PolComp}. This is presented in Figure \ref{fig:sens}(c).

In all three cases {\bf A1--A3}, the main result that staggered release leads to lower death rates for older groups and overall relative to the simultaneous release benchmark holds across a wide variety of choices for these parameters for which the correct values are less certain in the literature.

\subsection{Length of Initial Restrictions} 

In the main text, we used $d_b = 30$ days. 
Our simulations with $d_b=60$ and $d_b=90$ days show similar qualitative results as that for $d_b=30$ days presented in
Figure \ref{fig:comp3G}.

\begin{figure}[htb!]
\centering
\includegraphics[width=300pt]{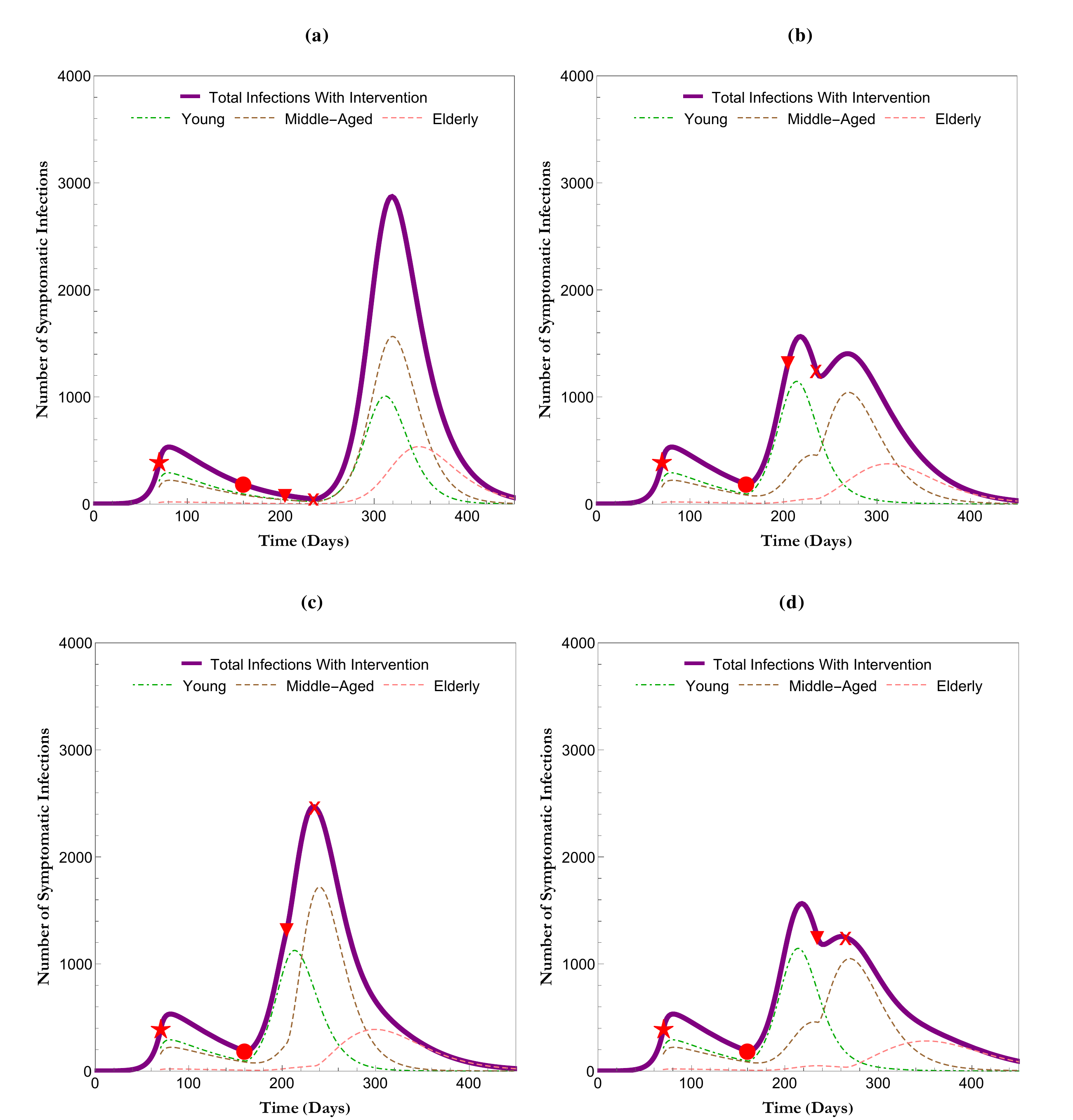}
\caption{
Similar to Figure \ref{fig:comp3G} but with $d_b=90$ for the initial duration ($d_b=30$ in Figure \ref{fig:comp3G}).  All other parameter values are the same as in Figure \ref{fig:comp3G}. The projected overall death rates are: (a) 1.38, (b) 1.26, (c) 1.29, and (d) 1.11, and these follow the pattern as the cases in Figure \ref{fig:comp3G}.
 }\label{fig:sens3}
\end{figure}

\end{appendix}

\end{document}